\documentclass[preprint
]{elsarticle}
\usepackage{graphicx}
\usepackage{lineno}

\begin{document}

\title{Observation of individual spin quantum transitions of a single antiproton}%

\begin{abstract}
We report on the detection of individual spin quantum transitions of a single trapped antiproton in a Penning trap. The spin-state determination, which is based on the unambiguous detection of axial frequency shifts in presence of a strong magnetic bottle, reaches a fidelity of 92.1$\%$. Spin-state initialization with $>99.9\%$ fidelity and an average initialization time of 24 min are demonstrated. This is a major step towards an antiproton magnetic moment measurement with a relative uncertainty on the part-per-billion level.
\end{abstract}

\author[riken,cern]{C. Smorra\corref{cor1}}
\ead{christian.smorra@cern.ch}
\author[riken]{A. Mooser}
\author[riken]{M. Besirli}
\author[riken,mpik]{M. Bohman}
\author[riken,iqo]{M. Borchert}
\author[mpik]{J. Harrington} 
\author[riken,ut]{T. Higuchi}
\author[riken,ut]{H. Nagahama}
\author[iphm]{G. L. Schneider}
\author[riken]{S. Sellner}
\author[riken,ut]{T. Tanaka}
\author[mpik]{K. Blaum}
\author[ut]{Y. Matsuda}
\author[iqo,ptb]{C. Ospelkaus}
\author[gsi]{W. Quint}
\author[iphm,him]{J. Walz}
\author[aplriken]{Y. Yamazaki}
\author[riken]{S. Ulmer}

\cortext[cor1]{Corresponding author}
\address[riken]{Ulmer Initiative Research Unit, RIKEN, 2-1 Hirosawa, Wako, Saitama 351-0198, Japan}
\address[cern]{CERN, CH-1211 Geneva 23, Switzerland}
\address[mpik]{Max-Planck-Institut f\"ur Kernphysik, Saupfercheckweg 1, D-69117 Heidelberg, Germany}
\address[iqo]{Institute of Quantum Optics, Leibniz Universit\"at Hannover, Welfengarten 1, D-30167 Hannover, Germany}
\address[ut]{Graduate School of Arts and Sciences, University of Tokyo, 3-8-1 Komaba, Meguro-ku, Tokyo 153-8902, Japan}
\address[iphm]{Institut f\"ur Physik, Johannes Gutenberg-Universit\"at Mainz, D-55099 Mainz, Germany}
\address[ptb]{Physikalisch-Technische Bundesanstalt, Bundesallee 100, D-38116 Braunschweig, Germany}
\address[gsi]{GSI-Helmholtzzentrum f\"ur Schwerionenforschung, D-64291 Darmstadt, Germany}
\address[uhd]{Ruprecht-Karls-Universit\"at Heidelberg, D-69047 Heidelberg, Germany}
\address[him]{Helmholtz-Institut Mainz, D-55099 Mainz, Germany}
\address[aplriken]{Atomic Physics Laboratory, RIKEN, 2-1 Hirosawa, Wako, Saitama 351-0198, Japan}
\date{November 30, 2017}%
\maketitle


Spectroscopy based on the observation of quantum transitions in specific systems enables sensitive measurements with highest resolution. For example, the observation of individual electron quantum transitions from a fluorescent to a dark electronic state in a single barium ion \cite{ToschekBaIon,DehmeltBaIon} has led to the development of first optical frequency standards \cite{Bergquist1986}. The fractional precision of optical clocks based on single-quantum transition readout schemes has advanced to the level of 10$^{-18}$ \cite{SingleIonClock}. Observations of single flux-quanta in superconductors provide sensitive magnetometers, represent accurate resistance standards, and measurements on quantized resistance in 2-dimensional electron gases led to precise measurements of the Planck constant \cite{h2011,h2012}. Experiments based on the detection of individual quantum transitions of single trapped electrons provide the most precise measurement of the fine-structure constant \cite{Hanneke2008}. Quantum-transition based spectroscopy of the magnetic anomalies $a_{e+/e-} = (g_{e+/e-}-2)/2$ of the electron and the positron in Penning traps led to a stringent test of the fundamental charge-parity-time (CPT) invariance \cite{Dehmelt,DehmeltCPT}, which is one of the cornerstones of the Standard Model of particle physics \cite{CPT}. In a recent experiment, the first observation of quantum transitions of a pure antimatter system has been made by inducing positron spin transitions in the ground state of antihydrogen \cite{ALPHA2012}.

All these experiments are based on quantum phenomena in electron/positron systems. Comparable observations in the proton/antiproton system require considerably higher experimental sensitivity caused by the different fundamental properties of the baryon system. Due to the 1836-fold higher masses and 658-fold smaller magnetic moments, the application of quantum-transition based spectroscopy schemes is more challenging compared to the electron/positron system \cite{DehmeltCSGE}. The observation of individual spin transitions of a single trapped proton has been recently demonstrated \cite{MooserPRL2013,JackPRL2013}. Based on this, we advanced to a high-precision measurement of the proton magnetic moment with 3.3$\cdot10^{-9}$ relative precision \cite{MooserNature2014}. 

Here, we report on the first non-destructive detection of individual spin transitions of a single antiproton using the continuous Stern-Gerlach effect \cite{DehmeltCSGE}. A magnetic bottle, $B_z=B_2(z^2-\rho^2/2)$, couples the antiproton spin to the axial motion of the particle. Thereby, the oscillation frequency, which is read out non-destructively \cite{Wine}, is modified depending on the spin state. This experiment was carried out in the analysis trap of the BASE Penning-trap system located at the antiproton decelerator facility (AD) of CERN \cite{SmorraEPJST2015}. The average fidelity of the spin-state identification is at 92.1$\%$ and spin-state initialization with $>99.9\%$ fidelity takes about 24 min. This allows the determination of the antiproton magnetic moment using high-precision measurement schemes \cite{haeffner2003double,MooserPLB2013} with the goal to reach a relative precision on the part-per-billion level \cite{MooserNature2014}. Thereby, we target a stringent test of CPT invariance with baryonic antimatter, potentially with more than a factor 100 improved relative precision compared to previous measurements of this quantity \cite{Jack2013Antiproton,HiroNC2016}. 


\begin{figure}[htb]
        \centerline{\includegraphics[width=0.47 \textwidth,keepaspectratio]{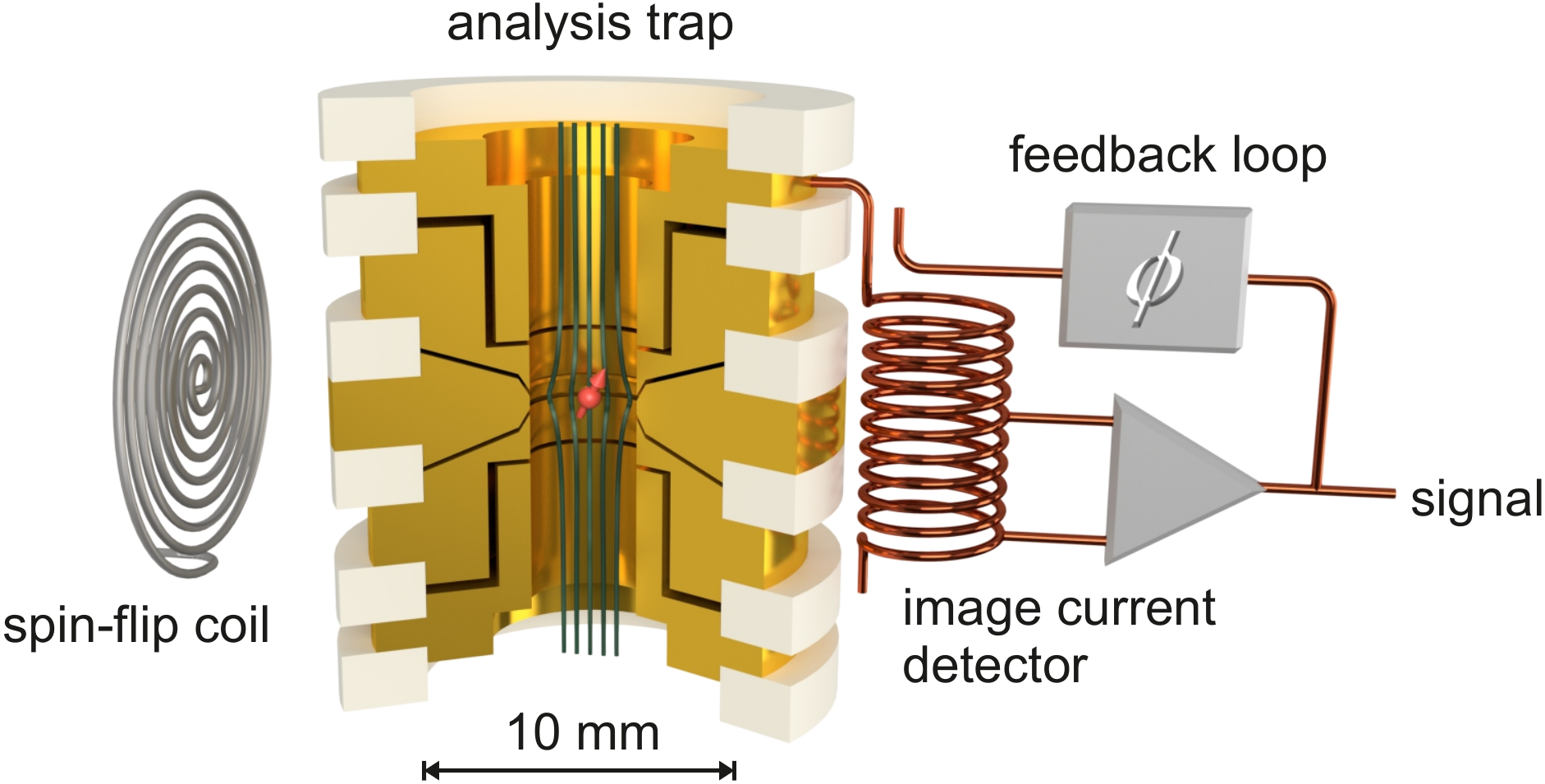}}
            \caption[Filter]{Experimental setup for the detection of single spin transitions in the BASE analysis trap. For details see text.} 
						\label{fig1}
    \end{figure}

Our apparatus, which is described in detail in ref.~\cite{SmorraEPJST2015}, consists of a cryogenic four Penning-trap system in the horizontal bore of a superconducting magnet with $B_0 = 1.945\,$T. The apparatus features a reservoir trap (RT), which serves as interface between the AD and the measurement traps and supplies single particles from the reservoir into the other traps when needed \cite{SmorraIJMS2015}. The precision trap (PT) and the cooling trap (CT) are required for the precision frequency measurements, and efficient cooling of the modified cyclotron mode, respectively. 

The measurements presented here were carried out in the analysis trap (AT), which is shown in Fig.$\,$\ref{fig1}. It is a 5-pole trap in orthogonal and compensated design and has 3.6 mm inner diameter \cite{CCRodegheri2012}. The central ring electrode is made out of a cobalt-iron alloy and generates the magnetic bottle $B_z=B_0 + B_2 (z^2-\rho^2/2)$ with $B_0=1.227\,$T and $B_2 = 272(15)\,$mT/mm$^2$. The other electrodes are made from oxygen-free electrolytic (OFE) copper. A superconducting image-current detection system is connected to one of the endcap electrodes to measure the axial frequency of the trapped antiproton \cite{Wine,Ulm,HiroRSI2016}. A feedback loop is implemented to apply feedback cooling to reduce the antiproton's axial oscillation amplitude \cite{DUrso2003}. Spin-transitions are induced by irradiating an oscillating magnetic field via a spin-flip coil placed in close vicinity to the trap electrodes.

The ideal Penning trap and the three harmonic oscillators composing a trapped particle's motion are described in ref.~\cite{Brown}. We denote the eigenfrequencies of the trapped antiproton as $\nu_z = 675\,$kHz, $\nu_+ = 18.7\,$MHz and $\nu_- = 12\,$kHz for the axial, the modified cyclotron and the magnetron modes, respectively.
To apply the continuous Stern-Gerlach effect, we use the magnetic bottle, which generates the magnetic potential $\Phi_{B,z} = - (\mu_+ + \mu_- + \mu_{\overline{p}}) B_z$. Here, $\mu_\pm = q/(2 m) L_{\pm}$ are the magnetic moments of the orbital angular momentum in the modified cyclotron and magnetron modes $L_{\pm}$, and $\mu_{\overline{p}}$ the spin magnetic moment. This causes an axial frequency shift $\Delta\nu_z$ depending on the quantum numbers of the radial modes $n_+, n_-$ and the spin quantum number $m_s$
\begin{eqnarray}
\Delta\nu_z = \frac{h \nu_+}{4 \pi^2 m_{\overline{p}} \nu_z} \frac{B_2}{B_0} 
\left( 
\left(n_+ + \frac{1}{2}\right) + 
\frac{\nu_-}{\nu_+}\left(n_- + \frac{1}{2}\right)+\frac{g_{\overline{p}}}{2}m_s \right),
\end{eqnarray}
where $h$ and $m_{\overline{p}}$ denote the Planck constant and the antiproton mass, respectively. The individual contributions to the frequency shift $\Delta\nu_z$ are expected to be $61\,$mHz, $39\,\mathrm{\mu}$Hz, and $172\,$mHz for single quantum transitions in the cyclotron mode, the magnetron mode and a spin flip, respectively. As a  result, spin transitions can be detected by observing changes of the axial oscillation frequency, given that the changes in the quantum numbers of the radial modes remain sufficiently small during axial frequency measurements. However, quantum number fluctuations in the radial modes driven by spurious voltage noise on the order of 100$\,$pV/$\sqrt{\mathrm{Hz}}$ on the trap electrodes constitute a major challenge for the spin-state identification. This voltage noise drives electric dipole transitions in the radial modes with a heating rate $\partial n_\pm/\partial t \propto |E_\pm|$ \cite{MooserPLB2013}, with $E_\pm$ being the energies in the radial modes. To minimize the heating rate, we reduce the cyclotron and magnetron amplitudes by resistive and sideband cooling, respectively, to a sub-thermal state with $(E_+ + |E_-|)/k_B <$ 100 mK. 
This cooling procedure is described in detail in ref.~\cite{MooserPLB2013}. 

\begin{figure}[htb]
        \centerline{\includegraphics[width=0.48 \textwidth,keepaspectratio]{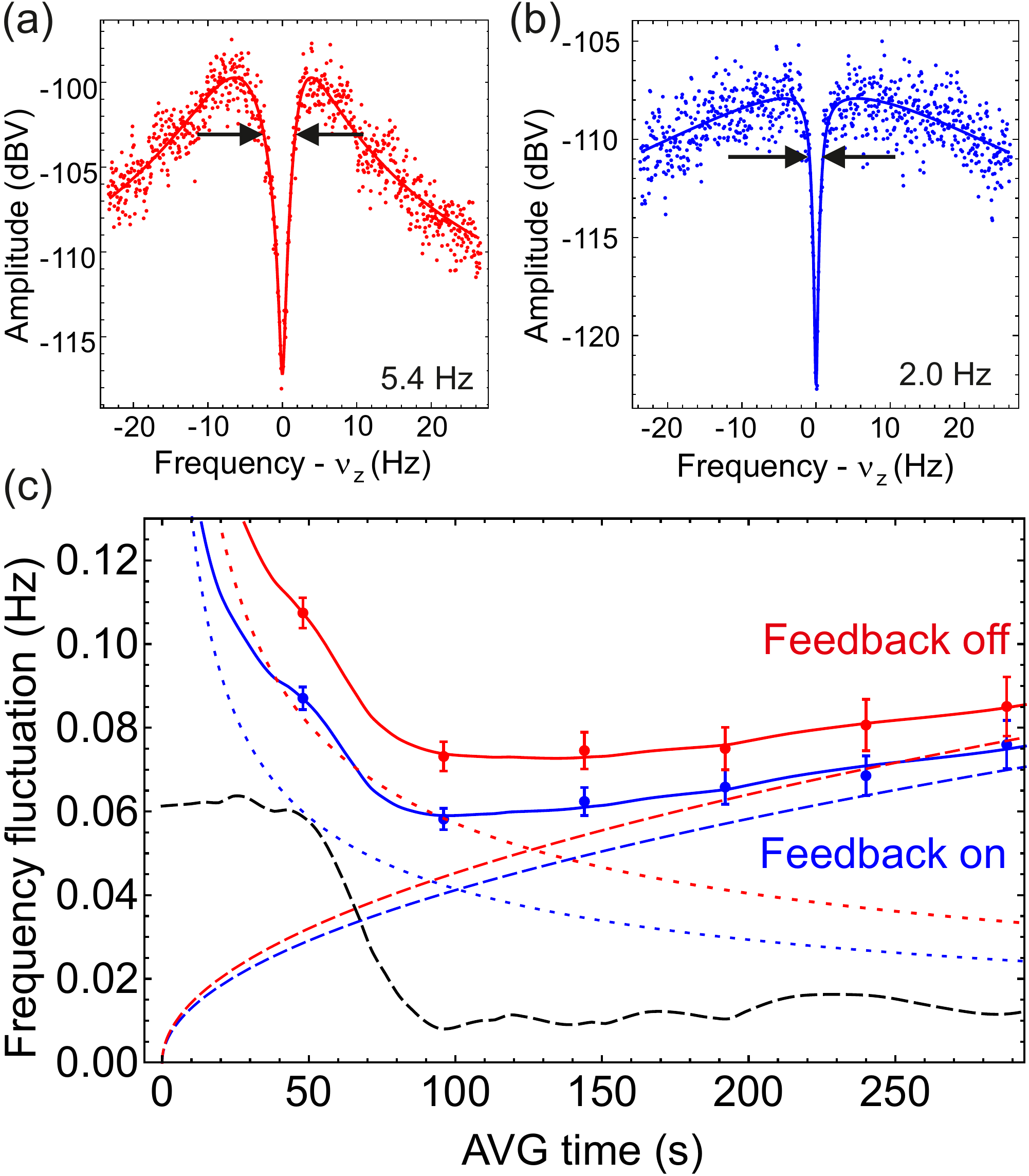}}
            \caption[Filter]{
						(a) and (b) show FFT spectra of the image-current signal of an antiproton dip without feedback (5.4 Hz line width at 5.7(4) K) and with axial feedback (2.0 Hz line width at 2.14(12) K), respectively. The points show the measured noise amplitudes and the solid line represents a fit of the theoretical line shape to the data. (c) Comparison of the axial frequency fluctuation $\Xi = \sigma(\Delta\nu_z)$ for the two measurement conditions as function of the averaging time. The measurements were performed at radial energies of $|E_-|/k_B < 7\,$mK and $E_+/k_B \approx 80\,$mK. 
						The points show the measured frequency fluctuation and the solid lines show fits of our fluctuation model to the data. The contributions of the individual components, the white noise and the random-walk noise, are shown as dotted lines and dashed lines, and a third component generated by the periodic magnetic field ramps of the AD is shown as black dashed line.} 
						\label{fig2a}
    \end{figure}

The axial frequencies are obtained from the FFT spectrum of the image-current signal from the trapped particle, as shown in Fig.$\,$\ref{fig2a} (a) and (b). In thermal equilibrium with the detection system, the antiproton appears as a `particle dip', which is a short of the resonator's Johnson-Nyquist noise at the antiproton's axial frequency. The line shape of the particle dip is well understood \cite{Wine}, and the axial frequencies are extracted from a least-squares fit to the data. The axial frequency stability determines the possibility of observing antiproton spin transitions. We perform subsequent measurements of $\nu_z$ and determine the frequency fluctuations $\Xi = \sigma(\left\{\nu_{k+1}-\nu_k\right\}_{k=1}^n)$ as a function of the averaging time $\tau$, as shown in Fig.$\,$\ref{fig2a} (c). The random-walk noise $\propto \tau^{1/2}$ (dashed line) increases $\Xi$ for averaging times longer than 100 s. At short averaging times, the white noise component $\propto \tau^{-1/2}$ (dotted line) mostly due to FFT averaging, and a component proportional to the Allan deviation of the magnetic field (black dashed line) caused by the periodic magnet ramps of the AD deceleration cycle impose limitations on $\Xi$. To reduce the white noise contribution, we apply feedback cooling in the axial mode \cite{CoolingMethods}. The detector signal is phase shifted by 180$^\circ$ and fed back to the trapped particle. The axial feedback grants a significant suppression of the axial frequency fluctuations, due to the reduced line width of our particle signal \cite{HiroRSI2016} and lower amplitude-dependent frequency shifts \cite{DUrso2003} by anharmonic contributions to the trapping potential. Using the feedback system, we have reached frequency fluctuations which are at the best conditions below 40 mHz at 96 s averaging time. The feedback system has been crucial for the detection of individual spin transitions with high fidelity, and allowed higher spin state detection fidelity than we reported for protons \cite{MooserPLB2013}. The cyclotron heating rates extracted from the random-walk component of the frequency fluctuations have reached comparable values in both experiments: $\partial n_+/\partial t =$ 0.035(4) K$^{-1}$ s$^{-1}$.

\begin{figure}[htb]
        \centerline{\includegraphics[width=0.48 \textwidth,keepaspectratio]{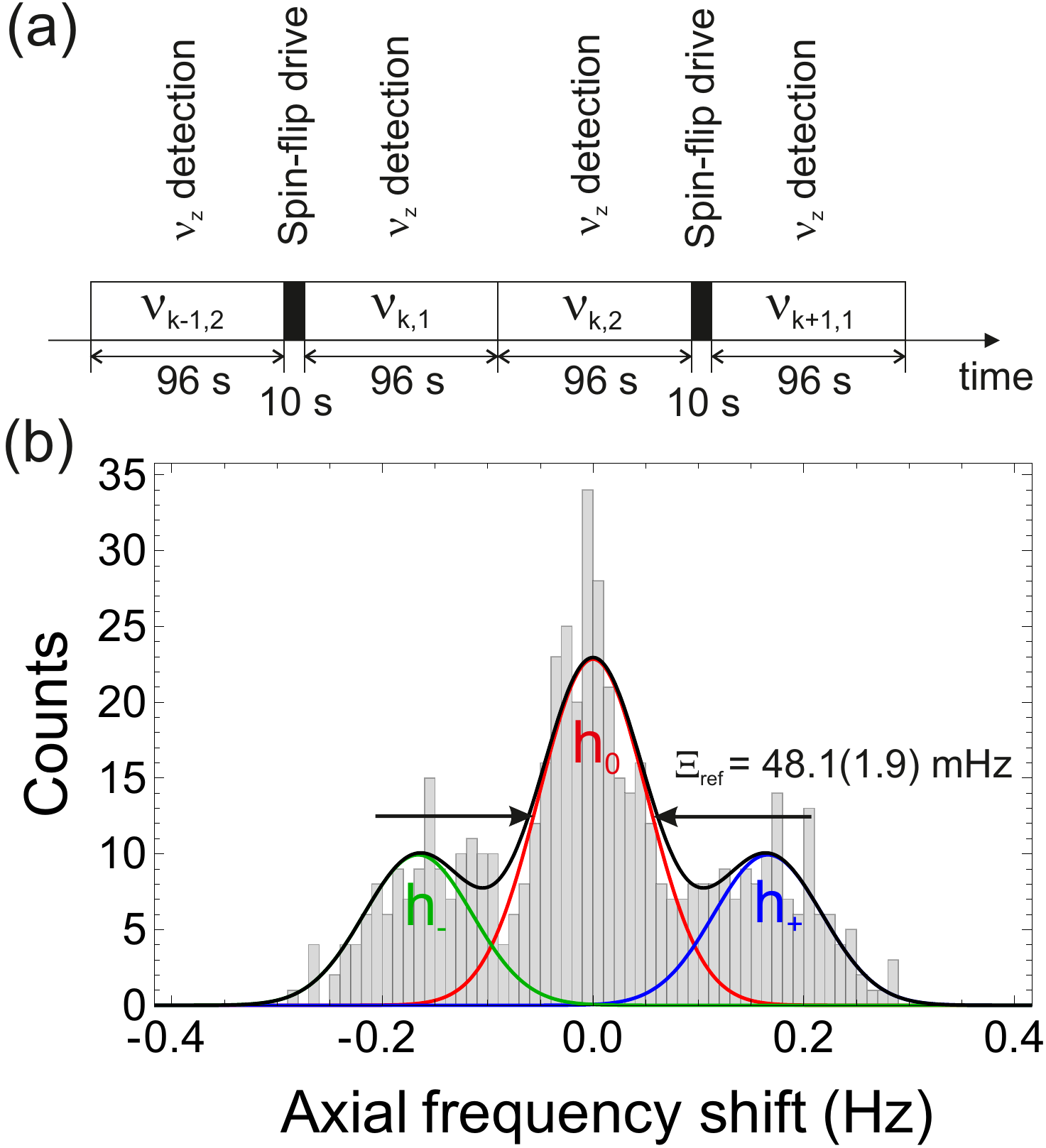}}
            \caption[Filter]{ (a) The measurement cycle for the spin transition detection is shown. (b) Histogram of axial frequency shifts for 96 s averaging time with resonant spin-flip drive at 52.3385 MHz. The black line shows the scaled probability density functions of this data with parameters determined from a likelihood analysis. We extract $\Xi_{\mathrm{ref}}=48.1(1.9)\,$mHz, $P_{\mathrm{SF}}=47.3(2.3)\,\%$, $\Delta\nu_{\mathrm{SF}}=166(4)\,$mHz and $B_2 = 262(6)\,$mT/mm$^2$, which is consistent with the value obtained from other measurements.} 
						\label{fig2}
    \end{figure}

To observe single antiproton spin transitions, we apply the measurement sequence shown in Fig.$\,$\ref{fig2} (a). One measurement cycle consists of two axial frequency measurements at 96$\,$s averaging time followed by a spin-flip drive. The axial frequency shifts $\nu_{k,2}-\nu_{k,1}$ characterize the axial frequency fluctuations, and the frequency differences $\nu_{k,1}-\nu_{k-1,2}$ are used to analyze the occurrence of spin transitions. Fig.$\,$\ref{fig2} (b) shows a histogram for 543 measurement cycles for the respective frequency shifts with resonant spin-flip drive. It is composed of three Gaussian distributions with the probability density functions (PDF) $h_0(\Delta,0,\Xi_{\mathrm{ref}}), h_-(\Delta,-\Delta\nu_{\mathrm{SF}},\Xi_{\mathrm{ref}})$ and $h_+(\Delta,+\Delta\nu_{\mathrm{SF}},\Xi_{\mathrm{ref}})$ describing the events with no spin transition, transition to the spin-down state and spin-up state, respectively. $\Delta$ is the frequency shift of the spin-flip drive, and the second and third parameter of the PDFs are the mean value and the standard deviation of the distribution, respectively, which we suppress in the following for a compact notation. 

The probability to observe a frequency shift $\Delta$ for an individual drive can be derived from the PDF
\begin{eqnarray}
p(\Delta,P_{\uparrow}) = P_\mathrm{SF} \left(P_{\uparrow} h_-(\Delta) + (1 - P_{\uparrow}) h_+(\Delta)\right)+ 
(1-P_\mathrm{SF}) h_0(\Delta),
\label{exhibit1}
\end{eqnarray}
where $P_{\uparrow}$ is the probability that the antiproton is initially in the spin-up state, and $P_\mathrm{SF}$ the spin-flip probability at given rf-drive parameters, which is usually optimized to 50$\%$ \cite{BrownGeoniumLineshape}. Under our experimental conditions it is possible to clearly distinguish the contributions of frequency shifts from the three distributions $h_0$ and $h_\pm$ in the data shown in Fig.$\,$\ref{fig2} (b). As the antiproton populates equally both spin states during the measurement sequence, we set $P_{\uparrow}=0.5$ in eq.~(\ref{exhibit1}) and extract the parameters of the PDFs from a likelihood analysis. From this we determine the mean frequency fluctuation $\Xi_{ref}=48.1(1.9)\,$mHz, the spin-flip probability $P_\mathrm{SF}=47.3(2.3)\,\%$, and the spin-flip frequency shift $\Delta\nu_{\mathrm{SF}}=166(4)\,$mHz. 

\begin{figure}[htb]
\centerline{\includegraphics[width=0.45 \textwidth,keepaspectratio]{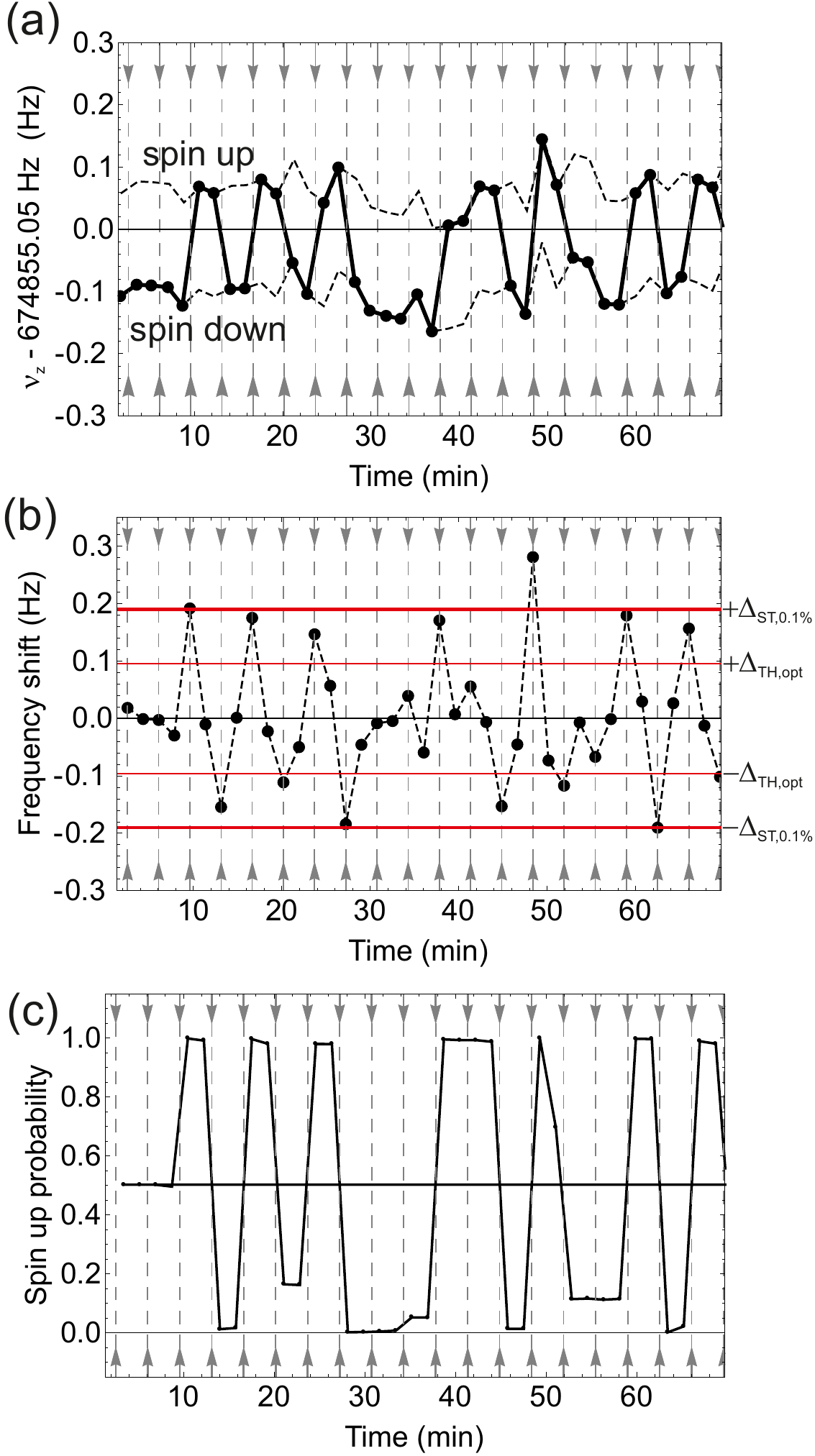}}
\caption[Filter]{Observation of single spin-transitions with antiprotons. (a) The points connected with the solid line show the measured axial frequencies with an offset of 674855.05$\,$Hz. The dashed lines allow the comparison of the measured frequency to the one which the particle would have in the opposite spin state. The gray arrows and dashed lines indicate the times of resonant spin-flip drives at 52.3385$\,$MHz. (b) The frequency shifts (points and dashed lines), the optimum threshold $\Delta_{\mathrm{TH,opt}}$=94$\,$mHz which minimizes the error rate $E_{\mathrm{TH}}$ (thin red line), and the threshold $\Delta_{f}$=190$\,$mHz where we obtain an initialization error rate $E_{i}$ of less than 0.1$\%$ (thick red line) are shown. (c) The propagation of the probability to be in spin state up using conditional probabilities (solid line) is shown. Details are given in the text.} 
\label{fig5}
\end{figure}

Fig.$\,$\ref{fig5} (a) shows the evolution of the axial frequency during a part of the measurement sequence in detail. 
The data already suggests that we can assign a spin-state to the antiproton for each measured axial frequency and that the spin up and spin down state can be clearly distinguished. For a comprehensive analysis, the corresponding axial frequency shifts shown in Fig.$\,$\ref{fig5} (b) are investigated. The simplest approach to identify individual spin transitions is to "digitize" the frequency shifts by assigning a spin-state to the particle after each measurement using a threshold method. Based on the threshold parameter $\Delta_{\mathrm{TH}}$ the following events are assigned to each frequency shift $\Delta_k=\nu_{k,1}-\nu_{k-1,2}$,
\begin{eqnarray}
\Delta_k > \Delta_{\mathrm{TH}}: \mathrm{Transition\ to\ spin\ up}\nonumber\\
- \Delta_{\mathrm{TH}} < \Delta_k < \Delta_{\mathrm{TH}}: \mathrm{No\ spin\ transition} 
\label{thcond} \\
\Delta_k < -\Delta_{\mathrm{TH}}: \mathrm{Transition\ to\ spin\ down}.\nonumber
\end{eqnarray}
After the observation of a frequency shift $\Delta_k>\Delta_{\mathrm{TH}}$ ($\Delta<-\Delta_{\mathrm{TH}}$) the spin state of the antiproton is assumed to be in the spin up (down) state. If $|\Delta_k|<\Delta_{\mathrm{TH}}$ the spin state remains unchanged and we assign the same spin state as after the last identified spin transition. To address the uncertainty of the spin-state assignment, we can determine the conditional probability for the particle to be in the spin-up state given the observation $\left\{\Delta_k\right\}_{k=1}^{n}$,
\begin{eqnarray}
P(\uparrow_n|\left\{\Delta_k\right\}_{k=1}^{n}) = \nonumber\\ 
\frac{h_0(\Delta_n) P(\uparrow_{n-1}|\left\{\Delta_k\right\}_{k=1}^{n-1})(1-P_\mathrm{SF})+h_+(\Delta_n) (1-P(\uparrow_{n-1}|\left\{\Delta_k\right\}_{k=1}^{n-1})) P_\mathrm{SF}}{p(\Delta_n,P(\uparrow_{n-1}|\left\{\Delta_k\right\}_{k=1}^{n-1}))}.
\end{eqnarray}
This recursive expression depends on all frequency shift measurements in the sequence $\left\{\Delta_k\right\}_{k=1}^{n}$ and is initialized using maximum ignorance $P(\uparrow_0) = 0.5$ as starting condition, before any frequency shifts $\Delta_k$ are measured. The solid lines in Fig.$\,$\ref{fig5} (c) shows the evolution of the spin-up probabilities during the measurement sequence shown in Fig.$\,$\ref{fig5} (a). This demonstrates that we can assign the spin state in most cases with low uncertainty. For about 2/3 of our data we have less than $5\%$ probability that the particle is not in the assigned spin state.

Mean error rates of our spin-state analysis can be calculated if the parameters $\Xi_{\mathrm{ref}}$, $\Delta\nu_{\mathrm{SF}}$ and $P_{\mathrm{SF}}$ are known.
To obtain a compact notation for the error rates, we define the following integrals over the distributions $h_0$ and $h_{+/-}$ shown in Fig.$\,$\ref{fig2} (b): 
\begin{eqnarray}
F_+ = P_{\mathrm{SF}} \int_{\Delta_{TH}}^{\infty} \partial\Delta\, h_+(\Delta)\\
E_- = P_{\mathrm{SF}} \int_{\Delta_{TH}}^{\infty} \partial\Delta\, h_-(\Delta)\\
E_0 = (1-P_{\mathrm{SF}}) \int_{\Delta_{TH}}^{\infty} \partial\Delta\, h_0(\Delta)\\
\tilde{E}=P_{\mathrm{SF}} \int_{-\Delta_{TH}}^{\Delta_{TH}} \partial\Delta\, h_+(\Delta)\\
\tilde{F}=(1-P_{\mathrm{SF}}) \int_{-\Delta_{TH}}^{\Delta_{TH}} \partial\Delta\, h_0(\Delta).
\end{eqnarray}

\begin{figure*}[htb]
        \centerline{\includegraphics[width=0.95 \textwidth,keepaspectratio]{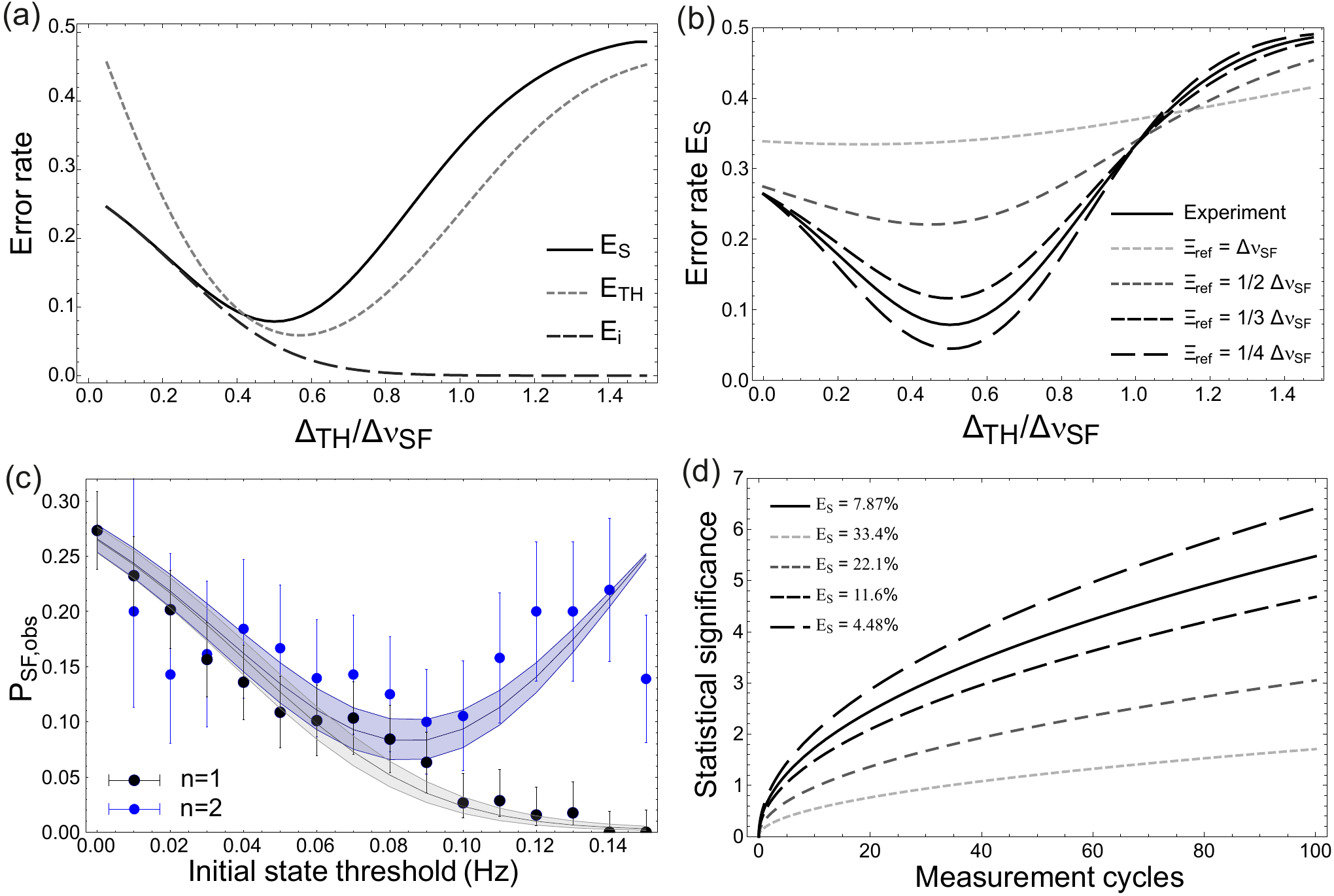}}
            \caption[Filter]{(a) The scaling of the error rates as function of the threshold parameter $\Delta_{\mathrm{TH}}$ is shown for the threshold method error rate $E_{\mathrm{TH}}$, for the initialization error rate $E_i$, and for the mean error rate of the spin-state assignment $E_S$. (b) The error rate $E_S$ as function of $\Delta_{\mathrm{TH}}$ is shown for different ratios $\Xi_{ref}/\Delta\nu_{\mathrm{SF}}$. (c) Observed spin-flip probabilities of simulated precision trap spin-flip drives with $P_{\mathrm{SF,PT}}$ = 0. A simulated drive was inserted in the measurement sequence for each observed frequency shift with $|\Delta_f| > 150\,$mHz. The observed spin-flip probability was evaluated with fixed threshold for the initial state, $\Delta_i$ = 150 mHz, and varying the threshold parameter for the final state $E_{f,n}$ for events requiring $n=1$ (black data points) and $n=2$ (blue data points) drives to identify the final state. The uncertainties of the data points originate from binomial or Poisson statistics. The results are compared to the theoretical expectations $P_{\mathrm{SF,obs}}=E_f (1-E_{i,n})+E_{i,n} (1-E_f)$ shown as solid lines. The shaded area represent the uncertainty of the theoretical prediction due to the uncertainties of the parameters $\Xi_{\mathrm{ref}}$, $P_{\mathrm{SF}}$ and $\Delta\nu_{\mathrm{SF}}$. (d) The statistical significance $s$ in standard deviations of distinguishing resonantly driven spin-flips of a saturated resonance in the PT from the background. The curve parameter is the error rate $E_S$. The parameters are corresponding to the minima of $E_S$ of the curves shown in Fig.$\,$\ref{fig3} (b). } 
						\label{fig3}
    \end{figure*}

For a single spin-flip drive, the probability that we interpret the information that we observed a frequency shift $\Delta$ in our axial frequency measurement sequence incorrectly, based the chosen threshold, is given by 
\begin{eqnarray}
E_{\mathrm{TH}} = P(|\Delta|<\Delta_{\mathrm{TH}}\,|\downarrow_0\uparrow_1\cup\uparrow_0\downarrow_1)+\nonumber\\
 P(\Delta >\Delta_{\mathrm{TH}}\,|\overline{\downarrow_0\uparrow_1}) + P(\Delta < -\Delta_{\mathrm{TH}}\,|\overline{\uparrow_0\downarrow_1})=\nonumber\\
\tilde{E} + 2 E_0 + E_-,
\end{eqnarray}
where the indices 0 and 1 indicate the spin state before and after the drive, respectively, and $\overline{\downarrow_0\uparrow_1}$ is the complementary event to $\downarrow_0\uparrow_1$. The first term is the fraction of occurring spin-transitions disregarded by the threshold, and the second and third term are the fraction of misidentified spin transitions due to axial frequency fluctuations. In reference to Fig.$\,$\ref{fig2} (b) these terms represent the overlaps of the distributions $h_0$ with $h_+$ and $h_-$. The scaling of $E_{\mathrm{TH}}$ as function of the threshold parameter $\Delta_{\mathrm{TH}}$ is shown in Fig.$\,$\ref{fig3} (a). $E_{\mathrm{TH}}$ can be minimized by chosing the optimum threshold $\Delta_{\mathrm{TH,opt}}=\Delta\nu_{\mathrm{SF}}/2\,(1+(2 \Xi_{ref}^2/\Delta\nu_{\mathrm{SF}}^2) \mathrm{ln}(2/P_{\mathrm{SF}}-2))$, which is for our experimental conditions at $\Delta_{\mathrm{TH,opt}} = 94\,$mHz and yields $E_{\mathrm{TH,opt}} = 5.8\,\%$. 

The error rate in the assignment of a spin state is different from $E_{\mathrm{TH}}$. We need to consider that this requires in general the observation of a spin transition in the measurement sequence, and that the spin state after the drive can be identified correctly even if we interpret the spin state based on $\Delta$ before the drive incorrectly. The spin-state assignment can require the application of several spin-flip attempts due to the incoherence caused by the interaction of the particle to the detection system. The coupling to the detector causes the amplitude of the axial motion to follow a Boltzmann distribution, which changes the average magnetic field experienced by the particle in the magnetic bottle. Under these conditions the maximum achievable spin-flip probability is $P_{\mathrm{SF,max}} = 0.5$ \cite{BrownGeoniumLineshape}. 

The initialization of the spin state in our sequence requires the observation of a spin transition $|\Delta| > \Delta_{\mathrm{TH}}$. We define the probability of assigning the wrong spin state after the observation of such an event as the initialization error rate $E_i$:
\begin{eqnarray}
E_{i} = \frac{P(\Delta>\Delta_{\mathrm{TH}}\,\cap\downarrow_n) + P(\Delta<-\Delta_{\mathrm{TH}}\,\cap\uparrow_n)}{P( \left|\Delta\right|>\Delta_{\mathrm{TH}})}  = \nonumber \\
\frac{E_0 + E_-}{2 E_0 + E_- + F_+},
\end{eqnarray}
where the denominator $2 E_0 + E_- + F_+ = P_{\mathrm{SF,obs}}$ is the observation probability of spin flips at a given threshold $\Delta_{\mathrm{TH}}$. $E_i$ can be reduced to an arbitrarily small value just by increasing $\Delta_{\mathrm{TH}}$. This allows to initialize a measurement sequence with a high fidelity as shown by the dashed black line in Fig.$\,$\ref{fig3} (a), however the number of observed spin-flips $P_{\mathrm{SF,obs}}$ decreases also rapidly when $\Delta_{\mathrm{TH}}$ exceeds $\Delta\nu_\mathrm{SF}$. Under our practically chosen experimental conditions, we use for this purpose a threshold of 190 mHz and achieve a fidelity of $(1-E_{i}) > 99.9 \%$. This level of initialization fidelity is higher than those reported for protons \cite{MooserPRL2013,JackPRL2013}. In our measurement sequence we observe a probability of 14.9(2)$\,\%$ for these events, corresponding to an average preparation time of 23.7(4) mins for spin-state identification at such high fidelity.

To calculate the mean error rate of the spin-state assignment, the error rates of spin states $E_{f,n}$ in the sequence which use the spin-state information from a transition after $n$ spin-flip attempts need to be determined. In this case, we have to consider the error rates of $n-1$ drives without observed spin transition in addition to $E_{i}$. If an odd number of errors occurs in the sequence of the $n$ last drives, the spin state is not identified correctly. $E_{f,n}$ is most simply expressed by the recursive formula:
\begin{eqnarray}
E_{f,n} = \frac{\tilde{F} E_{f,n-1} + \tilde{E} (1-E_{f,n-1})}{(1- P_{\mathrm{SF,obs}})},
\end{eqnarray}
where $E_{f,1}=E_{i}$. Compared to $E_{i}$, $E_{f,n}$ for $n>1$ increases for high thresholds since the amount of disregarded spin transitions for the drives with frequency shifts below the threshold increases. To define the average error rate of the spin-state assignment $E_S$, we weight the error rates $E_{f,n}$ with the probability of their occurrence:
\begin{eqnarray}
E_S = \sum_n (1 - P_{\mathrm{SF,obs}})^{n-1} P_{\mathrm{SF,obs}} E_n.
\end{eqnarray}
Note that $E_i$ and $E_{f,n}$ give also the error rates when we exchange initial and final states in the spin-state analysis, which is equivalent to a time reversal. This is in particular needed when the spin state of a sequence before the first drive needs to be determined. $E_S$ for our experimental conditions is shown in Fig.$\,$\ref{fig3} (a). For $\Xi_{\mathrm{ref}} \leq \Delta\nu_{\mathrm{SF}}$/3, $E_S$ is minimized for $\Delta_{\mathrm{TH}} \approx \Delta\nu_\mathrm{SF}/2$. The scaling of $E_S$ as function of $\Xi_{\mathrm{ref}}$ is shown in Fig.$\,$\ref{fig3} (b).

To measure the antiproton $g$-factor with high precision, we aim at the application of the double-trap measurement scheme \cite{haeffner2003double,MooserPLB2013}. In this method, the frequency measurements of the Larmor frequency $\nu_L$ and the cyclotron frequency $\nu_c$ ($g/2 = \nu_L/\nu_c$) are carried out in the precision trap (PT) with a homogeneous magnetic field \cite{MooserNature2014}. The measurement requires the detection of spin transitions driven in the precision trap by identification of the initial and final spin state in the analysis trap. To obtain a low number of incorrectly identified spin-flips we stop the sequence for the spin state determination of the initial state only after observing a frequency shift larger than the threshold $\Delta_i$ and obtain the initial state with a low error rate $E_i$. To determine the final state after the PT spin-flip attempt, we need to determine the initial state of the spin-flip sequence in the AT, which has the error-rate $E_{f,n}$ depending on the number of spin-flip attempts $n$ needed to observe a spin transition. For this purpose we use the threshold $\Delta_f$, which minimizes $E_S$. We obtain for $\Delta_f$=83$\,$mHz a mean error rate for the spin-state identification of $E_S=7.9\,\%$. Our antiproton apparatus reaches a lower error rate compared to our values reported for the proton, where we extract $E_S = 10.2\,\%$ based on the reported experiment parameters \cite{MooserPRL2013}.

The measurement quantity for the $g$-factor resonance is the spin-flip probability in the PT $P_{\mathrm{SF,PT}}$ as function of $\nu_L/\nu_c$. The error rates of the spin-state identification in the AT modify the observed spin-flip probability:
\begin{eqnarray}
P_{\mathrm{SF,PT,obs}} = P_{\mathrm{SF,PT}} (E_{f,n} E_i + (1-E_{f,n})(1-E_i)) +\nonumber\\ (1 - P_{\mathrm{SF,PT}}) (E_{f,n} (1-E_i) + (1-E_{f,n})E_i).
\label{eq14}
\end{eqnarray}
This relation can be verified by simulating a double-trap measurement based on our experimental data. For this purpose, we insert simulated PT spin-flip drives with $P_{\mathrm{SF,PT}}=0$ in our measurement sequence, i.e.~we assume that the particle was transported to the precision trap for a spin-flip trial and returned with its spin-state unchanged. For each of these simulated drives, we investigate if the initial and final spin state are identical. According to eq.~(\ref{eq14}), we expect to see a spin-flip probability of $P_{\mathrm{SF,PT,obs}}=(E_{f,n} (1-E_i) + (1-E_{f,n})E_i)$. This defines the background spin-flip rate of the double-trap $g$-factor resonance for off-resonant drives. The comparison of the observed spin-flip probabilities extracted from our experimental data and the calculated values for $P_{\mathrm{SF,PT,obs}}$ is shown in Fig.$\,$\ref{fig3} (c). Here, we inserted a simulated spin-flip drive after all frequency shifts with $|\Delta| > 150\,$mHz = $\Delta_i$ into the spin-flip sequence. The dependence of $P_{\mathrm{SF,PT,obs}}$ on the threshold for the analysis of the final state $\Delta_f \leq \Delta_i$ is shown for events requiring $n=1$ and $n=2$ spin-flip drives to define the spin state. Within the measurement uncertainties provided by the analyzed data, the measured spin-flip probabilities and the calculation are in good agreement. 

For $\Delta_f = 100\,$mHz, the simulated drives with $n=1$ and $n=2$ constitute 71$\%$ of the experimental data, and we obtain $P_{\mathrm{SF,off}}= 5.5^{+2.5}_{-1.9}\,\%$, which would constitute the background rate for the spin-flip detection in the PT under these conditions. The observed spin-flip probability on resonance for $P_{\mathrm{SF,PT}}=1/2$ is independent of the error rates: $P_{\mathrm{SF,on}} = 1/2$. The statistical significance of observing spin transitions in the PT is given by
\begin{eqnarray}
s = \frac{P_{\mathrm{SF,on}}-P_{\mathrm{SF,off}}}{\sqrt{\Delta P_{\mathrm{SF,on}}^2+\Delta P_{\mathrm{SF,off}}^2}} \nonumber\\
= \frac{1/2 - E_{S}}{\sqrt{\frac{1}{N_{on}}\frac{1}{4}+\frac{1}{N - N_{on}}E_{S} (1-E_{S})}},
\end{eqnarray}
where $P_{\mathrm{SF,on}}$ and $P_{\mathrm{SF,off}}$ are the observed spin-flip probabilities on and off resonance, respectively, $N$ the total number of spin-flip attempts, and $N_{on}$ the number of spin-flip attempts on resonance. To simplify the expression, we assumed that the error rate of the initial state $E_i \approx 0$ is negligibly small so that $P_{\mathrm{SF,off}} = E_S$, considering the contribution of all values of $n$. 

The scaling of $s$ with the number of spin-flip trials $N$ is shown in Fig.$\,$\ref{fig3} (d) after optimizing the ratio of $N_{on}/N$. For our experimental conditions, we can refute the zero hypothesis of observing spin flips caused entirely by the spin state error rates by 5 standard deviations with 85 data points. This performance of the spin state spectroscopy enables an antiproton double-trap $g$-factor measurement with high contrast and will allow to reach a relative precision on the part-per-billion level \cite{MooserNature2014}.

In conclusion, we have observed for the first time individual spin quantum transitions of a single trapped antiproton. This was achieved by using the continuous Stern-Gerlach effect in a Penning trap with a superimposed magnetic bottle of 272(15)$\,$mT/mm$^2$. In our current experiment, the axial frequency fluctuation of the antiproton in the magnetic bottle is at 48.1$\,$mHz for 96 s averaging time. Under these conditions, we have demonstrated that 92.1$\%$ of the spin states detected in our measurement sequence are identified correctly. In addition, a more conservative choice of our threshold parameter enables us to initialize the spin quantum state with a fidelity of 99.9$\%$ in a preparation time of 24 minutes, which increases the contrast of a double trap $g$-factor resonance further. These achievements constitute a major step towards a measurement of the antiproton magnetic moment with a fractional precision on the part-per-billion level which will provide one of the most stringent tests of charge-parity-time invariance in the baryon sector. 

We acknowledge financial support of RIKEN Initiative Research Unit Program, RIKEN President Funding, RIKEN Pioneering Project Funding, RIKEN FPR program, RIKEN JRA program, the Max-Planck Society, the CERN fellowship program, and the EU (ERC Advanced Grant No. 290870-MEFUCO). We acknowledge support from CERN, in particular from the AD operation team.


\begin{thebibliography}{}

\bibitem{ToschekBaIon} T. Sauter \textit{et al.}, Phys. Rev. Lett. \textbf{57}, 1696 (1986).
\bibitem{DehmeltBaIon} W. Nagourney \textit{et al.}, Phys. Rev. Lett. \textbf{56}, 2797 (1986).
\bibitem{Bergquist1986} J. C. Bergquist \textit{et al.}, Phys. Rev. Lett. \textbf{57}, 1699 (1986).
\bibitem{SingleIonClock} N. Huntemann \textit{et al.}, Phys. Rev. Lett. \textbf{116}, 063001 (2016).
\bibitem{h2011} A. Eichenberger \textit{et al.}, Metrologia \textbf{48}, 133-141 (2011).
\bibitem{h2012} I. A. Robinson, Metrologia \textbf{49}, 113-156 (2012).
\bibitem{Hanneke2008} D. Hanneke \textit{et al.}, Phys. Rev. Lett. \textbf{100}, 120801 (2008).
\bibitem{Dehmelt} R.S. Van Dyck \textit{et al.}, Phys. Rev. Lett. \textbf{59}, 26 (1987).
\bibitem{DehmeltCPT} H. Dehmelt \textit{et al.}, Phys. Rev. Lett. \textbf{83}, 4694 (1999).
\bibitem{CPT} G. L\"uders, Ann. Phys. \textbf{2}, 1-15 (1957).
\bibitem{ALPHA2012} C. Amole \textit{et al.}, Nature \textbf{483}, 439 (2012).
\bibitem{DehmeltCSGE} H. Dehmelt and P. Ekstr\"om, Bull. Am. Phys. Soc. \textbf{18}, 72 (1973).
\bibitem{MooserPRL2013} A. Mooser \textit{et al.}, Phys. Rev. Lett \textbf{110}, 140405 (2013).
\bibitem{JackPRL2013} J. DiSciacca \textit{et al.}, Phys. Rev. Lett \textbf{110}, 140406 (2013).
\bibitem{MooserNature2014} A. Mooser \textit{et al.}, Nature \textbf{509}, 596–599 (2014).
\bibitem{Wine} D. J. Wineland and H. G. Dehmelt, J. Appl. Phys. \textbf{46}, 919 (1975).
\bibitem{SmorraEPJST2015} C. Smorra \textit{et al.}, Eur. Phys. J. Special Topics \textbf{224}, 3055-3108 (2015).
\bibitem{haeffner2003double} H. H\"affner \textit{et al.}, Eur. Phys. J. D \textbf{22}, 163 (2003).
\bibitem{MooserPLB2013} A. Mooser \textit{et al.}, Phys. Lett. B \textbf{723}, 78-81 (2013).

\bibitem{Jack2013Antiproton} J. DiSciacca \textit{et al.}, Phys. Rev. Lett. \textbf{110}, 130801 (2013).
\bibitem{HiroNC2016} H. Nagahama \textit{et al.}, Nat. Comm. \textbf{8}, 14084 (2017).
\bibitem{SmorraIJMS2015} C. Smorra \emph{et al.}, Int. J. Mass Spectr. \textbf{389}, 10-13 (2015).
\bibitem{CCRodegheri2012} C. C. Rodegheri \textit{et al.}, New J. Phys. \textbf{14}, 063011 (2012).
\bibitem{Ulm} S. Ulmer \emph{et al.}, Rev. Sci. Inst. \textbf{80}, 123302 (2009).
\bibitem{HiroRSI2016} H. Nagahama \emph{et al.}, Rev. Sci. Inst., submitted (2016).
\bibitem{DUrso2003} B. D'Urso \textit{et al.}, Phys. Rev. Lett. \textbf{90}, 043001 (2003).
\bibitem{Brown} L. S. Brown and G. Gabrielse, Rev. Mod. Phys. \textbf{58}, 233 (1986).
\bibitem{CoolingMethods} W. M. Itano \emph{et al.}, Physica Scripta \textbf{T59}, 106-120 (1995).
\bibitem{BrownGeoniumLineshape} L. S. Brown, Ann. Phys. \textbf{159}, 62 (1985).





\end{thebibliography}
\end{document}